\documentclass[times,twocolumn]{aastex63}
\usepackage[utf8]{inputenc}
\usepackage[normalem]{ulem}
\usepackage{amsmath}
\usepackage{array,multirow,graphicx}

\newcommand{\HI}{H\,{\sc i}\ }

\begin{document}

\title{Characterizing Pulsar Distances Using \HI Kinematics}

\author{Steven Romero-Ruiz}
%\email{sromeror@caltech.edu}
\affiliation{Cahill Center for Astronomy and Astrophysics, California Institute of Technology, Pasadena, CA 91125, USA}

\author[0000-0002-4941-5333]{Stella Koch Ocker}
\email{socker@caltech.edu}
\affiliation{Cahill Center for Astronomy and Astrophysics, California Institute of Technology, Pasadena, CA 91125, USA}
\affiliation{Observatories of the Carnegie Institution for Science, Pasadena, CA 91101, USA}

\correspondingauthor{Stella Koch Ocker}

\begin{abstract}
    Distance measurements are fundamental to radio pulsars' use as astrophysical probes of General Relativity and the interstellar medium. One of the primary methods for determining pulsar distances is \HI kinematics, which leverages the radial velocities of \HI absorption and emission features detected along pulsar lines-of-sight. This method necessarily assumes a model for Galactic rotation, our knowledge of which continues to evolve in both accuracy and precision. In this research note, we derive kinematic distances for $66$ pulsars with archival \HI radial velocity measurements using a state-of-the-art Galactic rotation curve. The results and software are provided in an online repository. %Wherever possible, we tighten the inferred distance bounds by assessing expected distributions for pulsar luminosities and interstellar electron densities, as well as other pulsars with precisely determined parallax distances. 
    Our kinematic distances differ by $<1\sigma$ from published parallaxes for nearly all pulsars in the sample that have both types of distance measurement available. Comparison to the NE2025 Galactic electron density model shows general consistency between measured and predicted distances.
\end{abstract}

\keywords{Neutron stars -- Astrometry -- Galactic rotation -- Interstellar medium}

\section{Introduction}

Distance measurements are required for a wide range of radio pulsar applications, including tests of General Relativity \citep{kramer2006}, constraining neutron star formation and velocities \citep{hansen1997,verbunt2017}, and studies of the ionized interstellar medium (ISM; \citealt{ne20011,ymw16,ne2025}). However, less than $10\%$ of the $\sim4500$ known pulsars have empirical distance measurements (rather than distance estimates based on a Galactic electron density model). Nearly a third of published pulsar distances are derived using \HI kinematics, which combines a model for Galactic rotation with radial velocities of \HI absorption and emission features observed along pulsar lines-of-sight (LOSs; \citealt{frail1990,verbiest2012}). This method tends to deliver the least precision when compared to, e.g., parallaxes determined with Very Long Baseline Interferometry (VLBI) or distances determined by association with globular clusters \citep{deller2019,pan2021_globs}. Nonetheless, the sample of pulsar distances is limited enough that robust determination of \HI kinematic distances remains important.

Previous catalogs of \HI kinematic distances include those published by \cite{frail1990} and \cite{verbiest2012}, and were primarily based on \HI radial velocities measured towards very bright pulsars observed with the Parkes and Arecibo telescopes \citep[e.g.][]{ables1976,weisberg1987}. As laid out in these studies, the kinematic method relies on the detection of \HI features that place bounds on the pulsar distance: When \HI absorption is detected in the pulsar radio spectrum, it is attributable to gas in front of the pulsar; the velocity of the furthest absorption feature thus yields a lower limit on the pulsar distance. When strong \HI emission is detected beyond the furthest absorption feature in the pulsar spectrum, then an upper distance limit can be inferred (though in practice, the inferred upper limits depend on the observational sensitivity and the optical depth of \HI clouds; see \citealt{frail1990} for further discussion). In many cases, only a lower or upper distance bound can be inferred due to a lack of detected absorption or emission. 

The kinematic method also suffers from uncertainties due to peculiar motions, which are relevant to both the \HI gas and the assumed Galactic rotation curve. Previous pulsar kinematic distance catalogs adopted flat universal rotation curves \citep[e.g.][]{fich1989} with IAU constants, albeit with special consideration of peculiar motions in the direction of the Perseus arm \citep{frail1990,verbiest2012,jing2023}. Since the publication of these catalogs, Galactic rotation curve models have improved thanks to a significant increase in parallax and proper motion measurements of rotation curve tracers, including masers \citep{reid2014,reid2019} and classical Cepheids \citep{mroz2019}. 

This research note derives \HI kinematic distances for radio pulsars from archival radial velocity measurements based on the \cite{reid2019} Galactic rotation curve. Section~\ref{sec:obs} describes the sample of radial velocities used and  Section~\ref{sec:methods} explains how the kinematic distances are inferred. The resulting distances are discussed in Section~\ref{sec:results} and compared to the NE2025 Galactic electron density model's predictions. Conclusions are in Section~\ref{sec:conc}.

\section{The Sample}~\label{sec:obs}

For this analysis, we consider all radio pulsars (including radio-loud magnetars) with \HI spectral line constraints in the literature. The sample consists of 66 sightlines, drawn primarily from \cite{frail1990} and \cite{verbiest2012} and references therein.\footnote{ \cite{ables1976,weisberg1987,manchester81,koribalski95,weisberg95,gaensler99,strom2000,gaensler2003,ord2002,johnston2001,tian08,leahy2008,weisberg2008,minter2008}} More recent \HI measurements are drawn from \cite{liu2025L,jiang2025,jing2023,liu2021,vanleeuwen2015,kothes18,rudnitskii2014}.

Pulsars in the sample are bright (median $S_{1400\ {\rm GHz}} = 8$ mJy) and span a broad range in Galactic coordinates ($-116^\circ \leq l \leq 160^\circ$; $-9^\circ \leq b \leq 26^\circ$) and DM ($3.2 \leq {\rm DM} \leq 787$ pc cm$^{-3}$). Thirty-seven of the sources have a detection of both absorption and emission that give both a velocity lower and upper bound on distance; the rest either have only a single upper or lower velocity bound. Twenty-seven pulsars in the sample have a velocity bound at the tangent point.

%-- 76 pulsars w/ \HI absorption measurements from \cite{frail1990} and refs. therein (including many from \cite{ables1976,weisberg1987}, Manchester (1981) \\
%-- all bright (give flux density lower limit) pulsars covering broad range in DM (3-600) \\
%-- xx have detection of absorption and emission that give both a velocity lower and upper bound on distance \\
%-- yy have only a single upper or lower velocity bound; zz have a velocity bound at the tangent point \\
%-- all low latitude (less than 10), covering the entire range of longitude space \\

%-- in addition to the 76 \HI measurements from \cite{frail1990}, we consider 18 additional measurements considered by \cite{verbiest2012} towards radio-quiet neutron stars in supernova remnants, radio-bursting neutron stars, and pulsar wind nebulae (original references in table), and  %additional measurements for PSR 0438 from Wang et al., an \HI measurement toward the Kes 73 supernova remnant surrounding AXP 1E1841-045 (Tian and Leahy 2008), \HI detection in the pulsar wind nebula of J1930$+$1852 (Leahy et al. 2008), and an \HI absorption measurement in SNR CTB 80 surrounding PSR B1951$+$32 (Strom and Stappers 2000) -- last three of these were considered in \cite{verbiest2012}

\section{Methods}~\label{sec:methods}

\subsection{Galactic Rotation Curve}

An \HI cloud in circular motion around the center of the Galaxy has an observed radial velocity $V_r$ that is related to its angular velocity $\omega$ as
\begin{equation}
    V_r = (\omega-\omega_0)R_0{\rm sin}\ l\ {\rm cos}\ b
\end{equation}
where $R_0$ and $\omega_0$ are the Galactocentric radius and angular velocity of the Sun, respectively, and $(l,b)$ are the Galactic longitude and latitude of the cloud. The cloud's angular velocity can be related to its circular velocity $\Theta = R\omega$, where the Galactocentric radius $R$ of the cloud can be computed as
\begin{equation}
    R = (R_0^2 + d^2 - 2R_0d{\rm cos}\ l)^{1/2}
\end{equation}
for a Sun-cloud distance $d$ \citep{fich1989}. 

Galactic rotation curves give $\Theta(R)$, which combined with an observed $V_r$ yields an estimate of the distance $d$. We use the rotation curve model fit by \cite{reid2019}, hereafter R19, to a large sample of trigonometric parallaxes and proper motions for masers observed with Very Long Baseline Interferometry. R19 infer a distance to the Sun $R_0 = 8.15\pm0.15$ kpc and a circular velocity $\Theta_0 = 236\pm7$ km s$^{-1}$. The rotation curve is parameterized using the universal rotation curve model of \cite{persic96}. Our calculations are based on a Python version of the Fortran subroutine provided by R19 to calculate $\Theta(R)$. 

\begin{figure*}[ht!]
    \centering
    \includegraphics[width=0.95\textwidth]{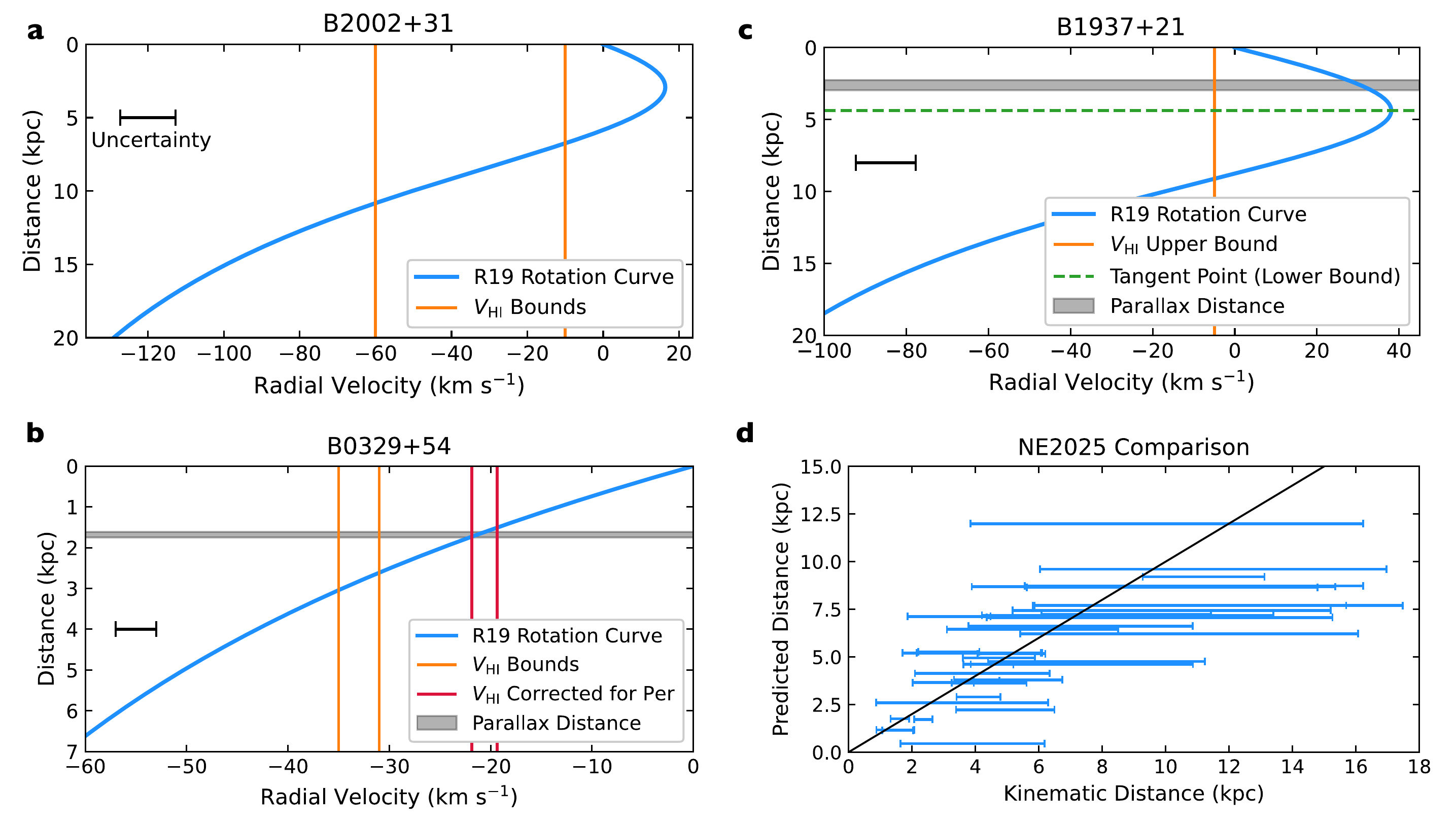}
    \caption{(a) Distance vs. radial velocity for PSR B2002$+$31. The R19 rotation curve model is shown in blue, and orange lines indicate the measured \HI velocity ($V_{\rm HI}$) bounds. A black errorbar indicates the magnitude of velocity uncertainty assumed due to random motions of \HI clouds. (b) Same as (a) for PSR B0329$+$54. Red lines indicate the velocity bounds corrected for peculiar motions in the Perseus arm. The independently measured parallax distance is shown in gray. (c) Same as (a,b) for PSR B1937$+$21. In this case the lower velocity bound is at the tangent point, indicated by the dashed green line. (d) Distances predicted by NE2025 vs. kinematic distances inferred in this work. }
    \label{fig:examples}
\end{figure*}

\subsection{Deriving Kinematic Distances from Radial Velocities}

Kinematic distances are determined based on a pulsar's known Galactic longitude and latitude, the assumed $\Theta(R)$ model, and reported \HI radial velocities inferred from the pulsar's radio spectrum. Radial velocity constraints fall into three main categories: (1) Lower and upper bounds; (2) single lower or upper bound; and (3) tangent point velocities. Tangent point velocities are those that lie at or above the maximum velocity expected for their sightlines. Figure~\ref{fig:examples} shows three pulsars illustrating these cases.

Sightlines through the Perseus arm, between longitudes $100^\circ \lesssim l \lesssim 150^\circ$, are well-known to be affected by significant peculiar motions \citep{belfort84,reid2019}. The standard correction is applied for these sightlines ($V_{r,{\rm corr}} = V_{r,{\rm obs}}/1.6$) \citep{frail1990,jing2023}. Throughout the analysis we adopt standard radial velocity uncertainties of $\pm 7$ km s$^{-1}$ to account for random cloud motions \citep{frail1990}, except in the direction of Perseus, for which this appears to overestimate the errors (see e.g. Figure~\ref{fig:examples}b) and we instead adopt 2 km s$^{-1}$ (the size of the smoothing kernel used in \citealt{weisberg2008}). 

\section{Analysis \& Results}~\label{sec:results}

\subsection{Revised Kinematic Distances}

The kinematic distances are provided in a Zenodo repository (\href{https://doi.org/10.5281/zenodo.19775798}{doi:10.5281/zenodo.19775798}). Previous studies used a variety of rotation curve models, all of which yield slightly different distance results than the R19 model applied here. On average, our distances differ by $\approx10\%$ from the distances reported in \cite{frail1990}, which are based on the \cite{fich1989} rotation curve. The most significant departures between our distances and those previously reported occur due to differences in rotation curves at $R<2.5$ kpc and $\approx 6-7$ kpc, where the circular velocity peaks in the R19 model. 

Of the 37 sources that have both a lower and upper bound on the distance, the vast majority of distance uncertainties are $>2$ kpc. Four pulsars have an \HI kinematic distance constraint with $<0.5$ kpc uncertainty: B0138$+$59, B0329$+$54, B0355$+$54, and B1855$+$09 (the first three are in the direction of Perseus, suggesting improvement to our uncertainty estimation may be needed); all but B0138$+$59 has a parallax distance measurement.

\subsection{Comparison to Parallax Distances}

Eight pulsars in the sample also have parallax distances.\footnote{J0332$+$5434, J0358$+$5413, J1820$-$0427, J1857$+$0943, J1932$+$1059, J1939$+$2134, J2018$+$2839, J2113$+$4644} In all but two cases, the parallax distances agree to within $1\sigma$ with the \HI kinematic distance; pulsars J2113$+$4644 and J2018$+$2839 both have parallaxes that are $>1\sigma$ smaller than the lower kinematic distance bounds, but they are still consistent to within $2\sigma$. None of these pulsars have parallax distances greater than the upper kinematic distance bounds. Two of the examples in Figure~\ref{fig:examples} show comparisons to independent parallax distances.

\subsection{Comparison to NE2025}

The kinematic distances are compared to NE2025 distance predictions \citep{ne2025} in Figure~\ref{fig:examples}. The predicted and measured distances generally agree to within the errors. We note that these \HI kinematic distances were not used to calibrate NE2025, although they are not a completely independent test of NE2025 because the previous model, NE2001, used kinematic distances based on the same \cite{frail1990} \HI measurements used here.

\section{Conclusions}~\label{sec:conc}

This research note provides an updated catalog of pulsar kinematic distances based on analysis of archival \HI radial velocity measurements with a uniformly applied, modern rotation curve model. There is significant room for improvement in the analysis, including deeper assessment of systematic uncertainties associated with the \HI measurements themselves and the corrections applied for peculiar and random motions. In the future we will explore methods to refine distance bounds through consideration of both pulsar and ISM properties, as well as technical advances in kinematic distance estimation based on \cite{wenger2018} and \cite{reid2022}. The distances inferred here are provided in an online repository (\href{https://doi.org/10.5281/zenodo.19775798}{doi:10.5281/zenodo.19775798}), along with a Python script for estimating pulsar kinematic distances from \HI measurements (\href{https://github.com/stella-ocker/psr-HI-kinematics}{github.com/stella-ocker/psr-HI-kinematics}).
%- ways to refine inferred distance bounds through consideration of pulsar and ISM properties -- note issues with using luminosity function as in verbiest 2012 \\

\acknowledgements S.K.O. is supported by the Brinson Foundation through the Brinson Prize Fellowship Program, and is a member of the NANOGrav Physics Frontiers Center (NSF award PHY-2020265). The authors acknowledge support from the Carnegie Institution for Science through the Carnegie Astrophysics Summer Student Internship.

\bibliography{master_bib}

\end{document}